\documentclass[epj]{webofc}
\woctitle{Physics Opportunities at an Electron-Ion Collider}
\usepackage[varg]{txfonts}   
\begin{document}
\title{Factorization theorem, gluon poles and new contributions in semi-inclusive processes}
%
%

\author{I.~V.~Anikin\inst{1}\fnsep\thanks{\email{anikin@theor.jinr.ru}} \and
        O.~V.~Teryaev\inst{1}\fnsep\thanks{\email{teryaev@theor.jinr.ru}}
}

\institute{Bogoliubov Laboratory of Theoretical Physics, JINR,
             141980 Dubna, Russia}

\abstract{
We discuss the role of gluon poles and the gauge invariance for the hadron tensors
of Drell-Yan and direct photon production processes with the transversely polarized hadron.
}
\maketitle
\section{Introduction}

We discuss the
direct photon production in two hadron collision where one hadron is transversely polarized.
We present the hadron tensor for this process and study
the effects which lead to the soft breaking of factorization
(or the universality breaking) through
the QED and QCD gauge invariance \cite{AT-DPP}.
As in \cite{AT-10}, the special role is played by the contour gauge for gluon fields.
We demonstrate that the prescriptions for the gluonic poles in the twist $3$ correlators
are dictated by the prescriptions in the corresponding hard parts.
We argue that the prescriptions in the gluonic pole contributions differ from each other
depending upon the initial or final state interactions of the diagrams under consideration.
The different prescriptions are needed to ensure the QCD gauge invariance.
We treat this situation as a breaking of the
universality condition resulting in factorization soft breaking.
The extra diagram contributions, which naively do not have an imaginary phase, is discussed in detail.
We also show that the new (``non-standard") terms
do contribute to the hadron tensor exactly as the ``standard" terms known previously.
This is exactly similar to the case of Drell-Yan process studied in \cite{AT-10}.

We also present the results of the
detailed analysis of hadron tensor in the Feynman gauge with the particular emphasis
on the QED gauge invariance \cite{AT-FG}.
We find that the QED gauge invariance can be maintained only by taking into account the non-standard diagram.
Moreover, the results in the Feynman and contour gauges coincide if the
gluon poles in the correlators $\langle\bar\psi\gamma_\perp A^+\psi\rangle$ are absent.
This is in agreement with the relation between gluon poles and the Sivers function which
corresponds to the "leading twist" Dirac matrix $\gamma^+$.
We confirm this important property by comparing the light-cone dynamics for different correlators.
As a result, we obtain the QED gauge invariant hadron tensor which
completely coincides with the expression obtained within
the light-cone contour gauge for gluons, see \cite{AT-10}.

\section{Direct Photon Production}

\begin{figure}[t]
\centerline{
\includegraphics[width=0.3\textwidth]{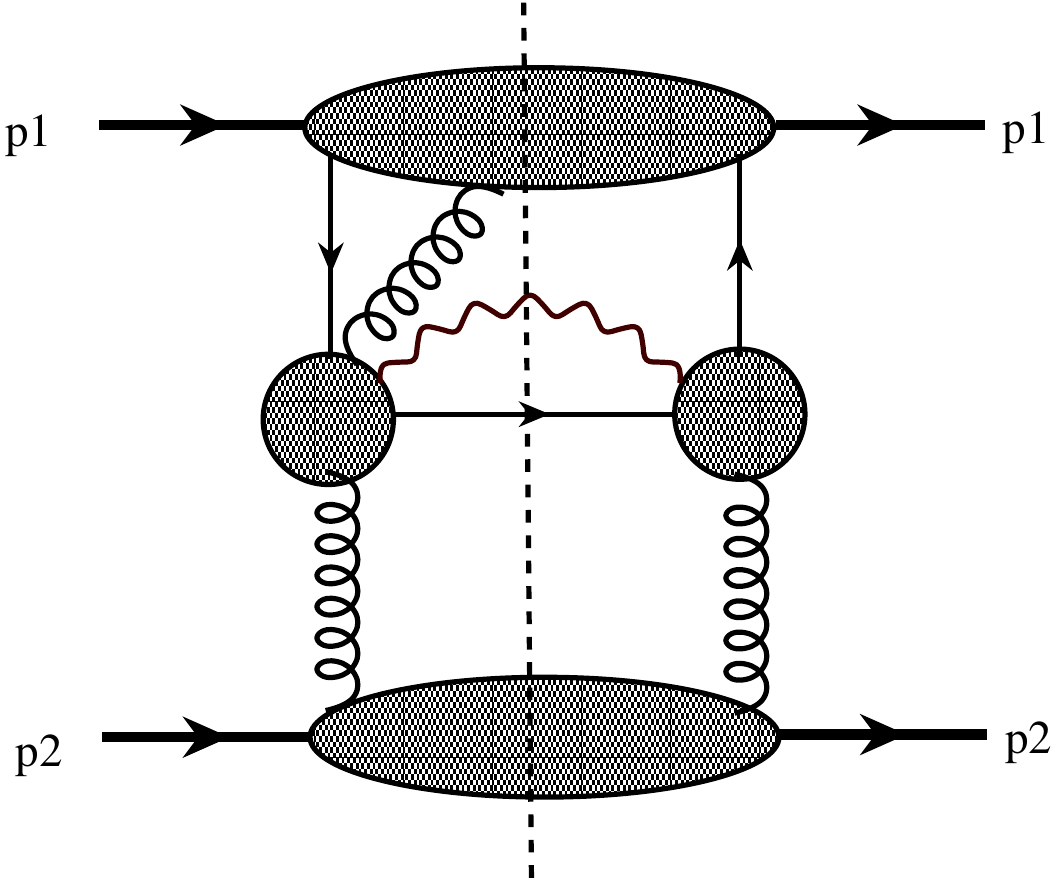}}
\caption{The Feynman diagram describing the hadron tensor of the
direct photon production.}
\label{Fig-DirPhot}
\end{figure}

We study the semi-inclusive process where the hadron with the transverse polarization
collides with the other unpolarized hadron to produce
the direct photon in the final state in:
\begin{eqnarray}
\label{process}
N^{(\uparrow\downarrow)}(p_1) + N(p_2) \to \gamma(q) + q(k) + X(P_X)\,.
\end{eqnarray}
For (\ref{process}) (also for the Drell-Yan process), the gluonic poles manifest.
We perform our calculations within a {\it collinear} factorization with kinematics defined as in
\cite{AT-DPP}.
So, we have the hadron tensor related to the corresponding asymmetry:
\begin{eqnarray}
\label{SSA}
d\sigma^\uparrow - d\sigma^\downarrow \sim {\cal W}=
\sum\limits_{i=1}^{2}\sum\limits_{j=1}^{8}
{\cal A}^{{\rm LO}}_i \ast {\cal B}^{{\rm NLO}}_j\,.
\end{eqnarray}
Here, we will mainly discuss the hadron tensor rather than the asymmetry
itself. So, the hadron tensor as an interference between the LO and NLO diagrams,
${\cal A}^{{\rm LO}}_i \ast {\cal B}^{{\rm NLO}}_j$, can be presented by Fig.~\ref{Fig-DirPhot}.

As a result of factorization, we should reach the factorized form for the considered subject:
\begin{eqnarray}
\label{Fac-DY}
\text{Hadron tensor} = \{ \text{Hard part (pQCD)}\} \otimes
\{ \text{Soft part (npQCD)} \}\,.
\end{eqnarray}
Usually, both the hard and soft parts, see (\ref{Fac-DY}), are independent of each other,
UV- and IR-renormalizable and, finally, various parton distributions, parametrizing the soft part,
have to manifest the universality property.
However, the hard and soft parts of the DPP hadron tensor are not fully independent each other \cite{AT-10}.
Actually, the DY hadron tensor has formally factorized with the mathematical convolution
and the twist-3 function $B^V(x_1,x_2)$ satisfies still the universality condition.
In contrast to the DY-process, the DPP tensor
will include the functions $B^V(x_1,x_2)$ that will not manifest the universality.

We now dwell on the QCD gauge invariance of the hadron tensor
for the direct photon production (DPP).
First of all, let us remind that
having used the {\it contour gauge conception} \cite{AT-DPP, ContourG},
one can check that the representation
\begin{eqnarray}
\label{B-plus}
B^V_{+}(x_1,x_2)= \frac{T(x_1,x_2)}{x_1-x_2 + i\epsilon}\,
\end{eqnarray}
belongs to the gauge defined by $[x,\,-\infty]=1$, while
the representation
\begin{eqnarray}
\label{B-minus}
B^V_{-}(x_1,x_2)=\frac{T(x_1,x_2)}{x_1-x_2 - i\epsilon}\,
\end{eqnarray}
corresponds to the gauge that defined by $[+\infty,\, x]=1$.
In both (\ref{B-plus}) and (\ref{B-minus}), the function $T(x_1,x_2)$ related to the
following prametrization:
\begin{eqnarray}
\label{parT}
\hspace{-0.7cm}\langle p_1, S^T | \bar\psi(\lambda_1 \tilde n)\, \gamma^+ \,
\tilde n_\nu G^{\nu\alpha}_T(\lambda_2\tilde n) \,\psi(0)
|S^T, p_1 \rangle=
\varepsilon^{\alpha + S^T -}\,(p_1p_2)\,
\int dx_1 dx_2 \, e^{i x_1\lambda_1+ i(x_2 - x_1)\lambda_2} \, T(x_1,x_2)\,.
\end{eqnarray}
Roughly speaking, it resembles the case where two different vectors have the same projection on the
certain direction. In this sense, the usual axil gauge $A^+=0$ can be understood as a ``projection" which
corresponds to two different ``vectors" represented by two different contour gauges.

Further, to check the gauge invariance,
we have to consider all typical diagrams, depicted in Fig.\ref{Fig-DirPhot}, that
correspond to the certain $\xi$-process (see, \cite{BogoShir}). Notice that, for the QCD gauge invariance, we have to assume
that all charged particles are on its mass-shells. That is, we will deal with only the physical gluons.

As a result of our calculation, the QCD gauge invariance (or the Ward identity) takes place provided only the
presence of the different complex prescriptions in gluonic poles dictated by the final or
initial state interactions:
\begin{eqnarray}
\label{QCD-illust}
  \begin{array}{rcl}
\hspace{-2cm}{\rm FSI}&\Rightarrow& \frac{1}{\ell^+ + i\epsilon} \Rightarrow {\rm gauge}\,\,\,[+\infty^-,\, z^- ]=1
\Rightarrow \frac{T(x_1,x_2)}{x_1-x_2 - i\epsilon}\,\\
\\
\hspace{-2cm}{\rm  ISI}&\Rightarrow& \frac{1}{-\ell^+ + i\epsilon} \Rightarrow {\rm gauge}\,\,\,[z^-,\, -\infty^-]=1
\Rightarrow \frac{T(x_1,x_2)}{x_1-x_2 +i\epsilon}\,\\
  \end{array}\,\,\, \Bigg\}
\Rightarrow \text{QCD Gauge Invariance}\,,
\end{eqnarray}
where $\text{FSI}$ and $\text{ISI}$ imply the final and initial state interactions, respectively.
We emphasize the principle differences between the considered case and the proof
of the QCD gauge invariance for the perturbative Compton scattering amplitude with the physical gluons in the
initial and final states. The latter does not need any external condition, like the presence of gluon poles.

We now calculate the full expression for the hadron tensor
which involves both the standard and new contributions to the gluon pole terms.
After computing the corresponding traces and performing simple algebra within the frame we are choosing,
it turns out that the only nonzero contributions to the hadron tensor come from the following terms
\begin{eqnarray}
\label{diaH1}
d{\cal W}(\text{H1})&=&\frac{d^3 \vec{q}}{(2\pi)^3 2 E} \int \frac{d^3 \vec{k}}{(2\pi)^3 2 \varepsilon}
\delta^{(2)}(\vec{\bf k}_\perp+\vec{\bf q}_\perp)
\, \text{C}_2\, \int dx_1 dy \delta(x_1-x_B)\,\delta(y-y_B)\,  {\cal F}^g(y)\, \times
\nonumber\\
&&\int dx_2\, \frac{2 S^2\, x_1 \, y^2}{[x_2 y S + i\epsilon][x_1 y S + i\epsilon]^2}\,
\frac{\varepsilon^{q_\perp + S_\perp -}}{p_1^+}\, B^{V}_-(x_1,x_2)\,,
\end{eqnarray}
\begin{eqnarray}
\label{diaH7}
d{\cal W}(\text{H7})&=&\frac{d^3 \vec{q}}{(2\pi)^3 2 E} \int \frac{d^3 \vec{k}}{(2\pi)^3 2 \varepsilon}
\delta^{(2)}(\vec{\bf k}_\perp+\vec{\bf q}_\perp)
\, \text{C}_1\, \int dx_1 dy \delta(x_1-x_B)\,\delta(y-y_B)\, {\cal F}^g(y)\, \times
\nonumber\\
&&\int dx_2\, \frac{(-2) S\, T\, x_1\, (y-3y_B)}{[x_2 T + i\epsilon][x_1 T + i\epsilon]^2}\,
\frac{\varepsilon^{q_\perp + S_\perp -}}{p_1^+}\, B^{V}_+(x_1,x_2)\,,
\end{eqnarray}
\begin{eqnarray}
\label{diaD4}
d{\cal W}(\text{D4})&=&\frac{d^3 \vec{q}}{(2\pi)^3 2 E} \int \frac{d^3 \vec{k}}{(2\pi)^3 2 \varepsilon}
\delta^{(1)}(\vec{\bf k}_\perp+\vec{\bf q}_\perp)
\, \text{C}_1\, \int dx_1 dy \delta(x_1-x_B)\,\delta(y-y_B)\,\frac{2}{S}  {\cal F}^g(y)\, \times
\nonumber\\
&&\frac{2 S^2 \,x_1\, (y-2y_B)}{[x_1 T + i\epsilon]^2}\,
\frac{\varepsilon^{q_\perp + S_\perp -}}{2x_1p_1^+ + i\epsilon}\, \int dx_2\,B^{V}_+(x_1,x_2)\,,
\end{eqnarray}
and
\begin{eqnarray}
\label{diaH10}
d{\cal W}(\text{H10})&=&\frac{d^3 \vec{q}}{(2\pi)^3 2 E} \int \frac{d^3 \vec{k}}{(2\pi)^3 2 \varepsilon}
\delta^{(2)}(\vec{\bf k}_\perp+\vec{\bf q}_\perp)
\,\text{C}_3\, \int dx_1 dy \delta(x_1-x_B)\,\delta(y-y_B)\, {\cal F}^g(y)\, \times
\nonumber\\
&&\int dx_2\, \frac{2 T (x_1-x_2) (2 T+S y)}{[x_1 T + i\epsilon][x_2 T + i\epsilon][(x_1-x_2) y S + i\epsilon]}\,
\frac{\varepsilon^{q_\perp + S_\perp -}}{p_1^+}\, B^{V}_+(x_1,x_2)\,.
\end{eqnarray}
Here, $\text{C}_1=C_F^2 N_c$, $\text{C}_2=-C_F/2$, $\text{C}_3=C_F\, N_c \,C_A/2$.
The other diagram contributions disappear owing to the following reasons: (i) the $\gamma$-algebra gives
$(\gamma^-)^2=0$; (ii) the common pre-factor $T+yS$ goes to zero,
(iii) cancelation between different diagrams.

Analysing the results for the terms H1, H7, D4 and H10 (see Eqns. (\ref{diaH1})--(\ref{diaH10})),
we can see that
\begin{eqnarray}
\label{Fac2}
d{\cal W}(\text{H1}) + d{\cal W}(\text{H7}) + d{\cal W}(\text{D4}) =
d{\cal W}(\text{H10})\,.
\end{eqnarray}
In other words, as similar to the Drell-Yan process, the new (``non-standard") contributions
generated by the terms H1, H7 and D4 result again in the factor of $2$ compared to the
``standard" term H10 contribution to the corresponding hadron tensor.

\section{Drell-Yan process}

\begin{figure}[t]
\centerline{\includegraphics[width=0.3\textwidth]{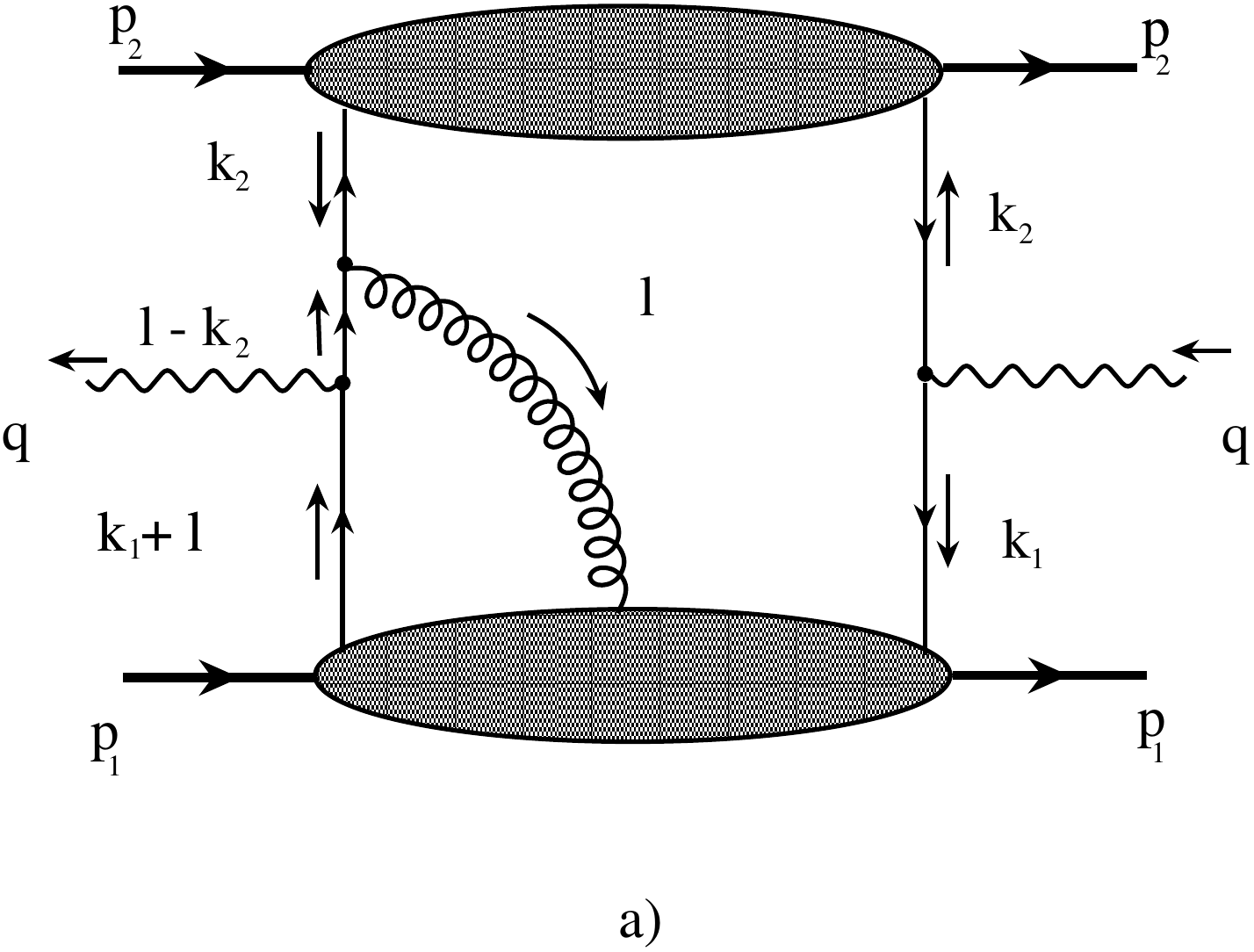}
\hspace{1.cm}\includegraphics[width=0.3\textwidth]{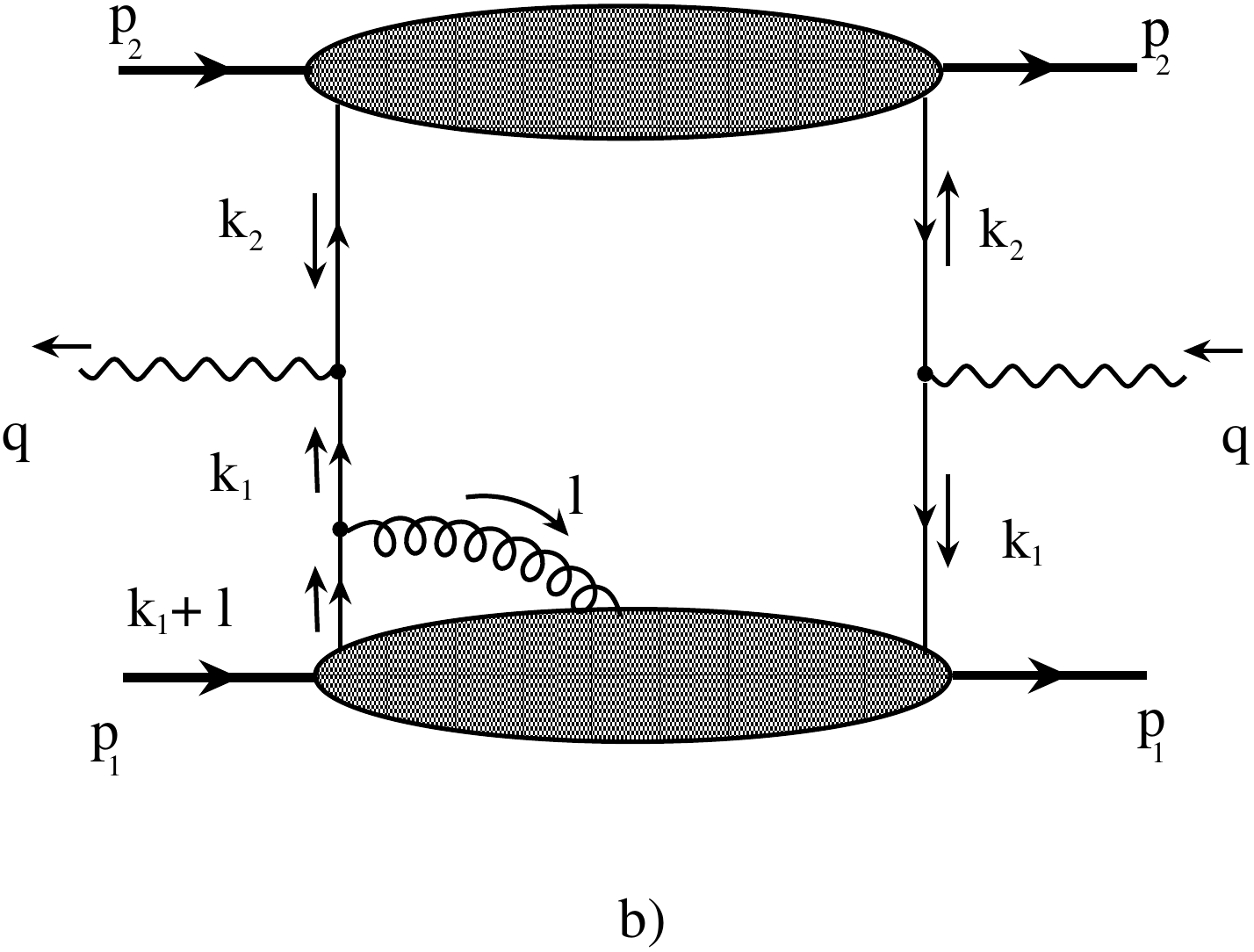}}
\caption{The Feynman diagrams which contribute to the polarized Drell-Yan hadron tensor.}
\label{Fig-DY}
\end{figure}

We now discuss the hadron tensor which contributes to the single spin
(left-right) asymmetry
measured in the Drell-Yan process with the transversely polarized nucleon (see Fig.~\ref{Fig-DY}):
\begin{eqnarray}
N^{(\uparrow\downarrow)}(p_1) + N(p_2) \to \gamma^*(q) + X(P_X)
\to\ell(l_1) + \bar\ell(l_2) + X(P_X).
\end{eqnarray}
Here, the virtual photon producing the lepton pair ($l_1+l_2=q$) has a large mass squared
($q^2=Q^2$)
while the transverse momenta are small and integrated out.
The left-right asymmetry means that the transverse momenta
of the leptons are correlated with the direction
$\textbf{S}\times \textbf{e}_z$ where $S_\mu$ implies the
transverse polarization vector of the nucleon while $\textbf{e}_z$ is a beam direction \cite{Barone}.
We perform our calculations within a {\it collinear} factorization with kinematics as in \cite{AT-10}.

We work within the Feynman gauge for gluons.
After calculation of all relevant traces in the factorized hadron tensor and after some algebra, we arrive at the following
contributions for the unintegrated hadron tensor (which involves all relevant contributions except the mirror ones):
the standard diagram depicted in Fig.~\ref{Fig-DY}(a) gives us
\begin{eqnarray}
\label{DY-St}
&&\overline{\cal W}^{(\text{Stand.})}_{\mu\nu}
 + \overline{\cal W}^{(\text{Stand.},\,\partial_\perp)}_{\mu\nu}=
 \bar q(y)\,
\Bigg\{
- \frac{p_{1\,\mu}}{y}\,
\varepsilon_{\nu S^T -p_2}\, \int dx_2 \frac{x_1-x_2}{x_1-x_2+i\epsilon} B^{(1)}(x_1,x_2) -
\\
&&
\Big[ \frac{p_{2\,\nu}}{x_1} \varepsilon_{\mu S^T - p_2} + \frac{p_{2\,\mu}}{x_1} \varepsilon_{\nu S^T - p_2} \Big]
x_1\int dx_2 \frac{B^{(2)}(x_1,x_2)}{x_1-x_2+i\epsilon}
+ \frac{p_{1\,\mu}}{y} \,
\varepsilon_{\nu S^T - p_2}\, \int dx_2 \frac{B^{(\perp)}(x_1,x_2)}{x_1-x_2+i\epsilon}
\Bigg\}\,,
\nonumber
\end{eqnarray}
while the non-standard diagram presented in Fig.~\ref{Fig-DY}(b)
contributes as
\begin{eqnarray}
\label{DY-NonSt}
&&\overline{\cal W}^{(\text{Non-stand.})}_{\mu\nu}=
\bar q(y)
\frac{p_{2\,\mu}}{x_1}
\varepsilon_{\nu S^T -p_2}
\int dx_2 \Big\{ B^{(1)}(x_1,x_2) +
B^{(2)}(x_1,x_2)
\Big\}.
\end{eqnarray}
In Eqns.~(\ref{DY-St}) and (\ref{DY-NonSt}), all indices have to be treated as contravariant ones independently of
the real position in the formulae.
We also introduce the shorthand notation:
$\varepsilon_{A B C D}= \varepsilon_{\mu_1 \mu_2 \mu_3 \mu_4} A_{\mu_1} B_{\mu_2} C_{\mu_3} D_{\mu_4}$.
Moreover, the parametrizing functions are associated with the following correlators:
\begin{eqnarray}
\label{ParFunB1}
&&B^{(1)}(x_1,x_2)=\frac{T(x_1,x_2)}{x_1-x_2+i\varepsilon} \Longleftarrow
{\cal F}_2\Big[\langle p_1, S^T| \bar\psi(\eta_1)\, \gamma^+ \, A^T(z)\, \psi(0) | S^T,p_1 \rangle \Big]\,,
\\
\label{ParFunB2}
&& B^{(2)}(x_1,x_2) \Longleftarrow
{\cal F}_2\Big[\langle p_1, S^T| \bar\psi(\eta_1)\, \gamma^\perp \, A^+(z)\, \psi(0) | S^T,p_1 \rangle \Big]\,,
\\
\label{ParFunBperp}
&& B^{(\perp)}(x_1,x_2) \Longleftarrow
{\cal F}_2\Big[\langle p_1, S^T| \bar\psi(\eta_1)\, \gamma^+\, \big(\partial^\perp\, \, A^+(z)\big)\, \psi(0) | S^T,p_1 \rangle \Big]\,,
\end{eqnarray}
where ${\cal F}_2$ denotes the corresponding Fourier transformation.
Summing up all contributions from the standard and non-standard diagrams, we finally obtain
the expression for the unitegrated hadron tensor. We have
\begin{eqnarray}
\label{DY-ht-1}
&&\overline{\cal W}_{\mu\nu}=
\overline{\cal W}^{(\text{Stand.})}_{\mu\nu} + \overline{\cal W}^{(\text{Stand.},\,\partial_\perp)}_{\mu\nu}+
\overline{\cal W}^{(\text{Non-stand.})}_{\mu\nu}=
\\
&&
\bar q(y)\,
\Bigg\{ \Big[ \frac{p_{2\,\mu}}{x_1} - \frac{p_{1\,\mu}}{y} \Big] \,
\varepsilon_{\nu S^T -p_2}\, \int dx_2 B^{(1)}(x_1,x_2) +  \frac{p_{2\,\mu}}{x_1} \,
\varepsilon_{\nu S^T - p_2}\, \int dx_2 B^{(2)}(x_1,x_2) -
\nonumber\\
&&
\Big[ \frac{p_{2\,\nu}}{x_1} \varepsilon_{\mu S^T - p_2} + \frac{p_{2\,\mu}}{x_1} \varepsilon_{\nu S^T - p_2} \Big]
x_1\int dx_2 \frac{B^{(2)}(x_1,x_2)}{x_1-x_2+i\epsilon} +
 \frac{p_{1\,\mu}}{y} \,
\varepsilon_{\nu S^T - p_2}\, \int dx_2 \frac{B^{(\perp)}(x_1,x_2)}{x_1-x_2+i\epsilon}
\Bigg\}\,,
\nonumber
\end{eqnarray}
Notice that the first term in Eqn.~(\ref{DY-ht-1}) coincides with the hadron tensor calculated within the light-cone gauge
$A^+=0$.

Let us now discuss the QED gauge invariance of the hadron tensor.
From Eqn.~(\ref{DY-ht-1}),
we can see that the QED gauge invariant combination is
\begin{eqnarray}
\label{GI-comb}
&&{\cal T}_{\mu\nu}=\Big[ p_{2\,\mu}/x_1 - p_{1\,\mu}/y \Big] \,
\varepsilon_{\nu S^T -p_2},\,
\quad
q_\mu {\cal T}_{\mu\nu} = q_\nu {\cal T}_{\mu\nu}=0.
\end{eqnarray}
At the same time, there is a single term with $p_{2\,\nu}$ which does not have
a counterpart to construct the gauge-invariant combination
$
p_{2\,\mu}/x_1 - p_{1\,\mu}/y
$.
Therefore, the second term in  Eqn.~(\ref{DY-St}) should be equal to zero.
This also leads to nullification of the second term in Eqn.~(\ref{DY-NonSt}).
Now, to get the QED gauge invariant combination (see (\ref{GI-comb})) one has no the other way rather than
to combine the first terms in Eqns.~(\ref{DY-St}) and (\ref{DY-NonSt}) which justifies the treatment of gluon pole in $B^{(1)}(x_1,x_2)$
using the complex prescription.
In addition, we have to conclude that the only remaining third term in (\ref{DY-St}) should not contribute to SSA.

Moreover, we can conclude that, in the case with the substantial transverse component of the momentum,
there are no any sources for the gluon pole at $x_1=x_2$. As a result, the function  $B^{(2)}(x_1,x_2)$ has no gluon poles and,
due to T-invariance, $B^{(2)}(x_1,x_2)= - B^{(2)}(x_2,x_1)$, obeys $B^{(2)}(x,x) = 0$.
On the other hand, if we have $\gamma^+$ in the correlator, see (\ref{ParFunB1}), the transverse components of gluon momentum
are not substantial and can be neglected. That ensures the existence of the gluon poles for the function $B^{(1)}(x_1,x_2)$.
This corresponds to the fact that the Sivers function, being related to gluon poles, contains the "leading twist" projector $\gamma^+$.
So, we may conclude that in the Feynman gauge the structure $\gamma^+ (\partial^\perp A^+)$ does not produce the imaginary part and SSA as well.

Working within the Feynman gauge, we derive the
QED gauge invariant (unitegrated) hadron tensor for the polarized DY process:
\begin{eqnarray}
\label{DY-ht-GI}
\overline{\cal W}^{\text{GI}}_{\mu\nu}= \bar q(y)
\Big[ p_{2\,\mu}/x_1 - p_{1\,\mu}/y \Big]
\varepsilon_{\nu S^T -p_2} \hspace{-1.5mm}\int dx_2 B^{(1)}(x_1,x_2).
\end{eqnarray}
Further, after calculation the imaginary part (addition of the mirror contributions),
and after integration over $x_1$ and $y$,
the QED gauge invariant hadron tensor takes the form
\begin{eqnarray}
\label{DY-ht-GI-2}
W^{\text{GI}}_{\mu\nu}= \bar q(y_B)\,
\Big[ p_{2\,\mu}/x_B - p_{1\,\mu}/y_B \Big] \,
\varepsilon_{\nu S^T -p_2}\,  T(x_B ,x_B)\,.
\end{eqnarray}
This expression fully coincides with the hadron tensor which
has been derived within the light-cone gauge for gluons.
Moreover, the factor of $2$ in the hadron tensor,
we found within the axial-type gauge \cite{AT-10}, is still present in the frame
of the Feynman gauge.

\section{Conclusions}

We explore the QCD gauge invariance of
the hadron tensor for the direct photon production in two
hadron collision where one of hadrons is transversely polarized.
We argue the effects which lead to the soft breaking of factorization
by inspection of the QCD gauge invariance.
We demonstrate that
the initial or final state interactions in diagrams define
the different prescriptions in the gluonic poles.
Moreover, the different prescriptions are needed to ensure the QCD gauge invariance.
This situation can be treated as a soft breaking of the
universality condition resulting in factorization breaking.
We find that the ``non-standard" new terms,
which exist in the case of the complex twist-$3$ $B^V$-function with the corresponding prescriptions,
do contribute to the hadron tensor exactly as the ``standard" term known previously.
This is  another important result of our work.
We also observe that this is exactly similar to the case of Drell-Yan process studied in \cite{AT-10}.

We argue the absence of gluon poles
in the correlators $\langle\bar\psi\gamma_\perp A^+\psi\rangle$
based on the light-cone dynamics.
At the same time, we also show that
the Lorentz and QED gauge invariance of the hadron tensor calculated within the Feynman gauge
requires the function $B^{(2)}(x_1,x_2)$ to be without any gluon poles.
This property seems to be natural from the point of view of gluon poles relation \cite{Boer:2003cm} to Sivers functions
as the latter is related to the projection $\gamma^+$. As for the function  $B^{(\perp)}(x_1,x_2)$,  the transverse derivative
of Sivers function resulting from taking its moments, may act on both integrand and boundary value. Our result suggests that only the
action on the boundary value, related to  $B^{(1)}(x_1,x_2)$ should produce SSA. It is not unnatural bearing in mind that the integrand
differentiation is present even for simple straight-line contours which are not producing SSA.

\end{document}